
\documentclass[aps,prl,twocolumn]{revtex4-2}

\usepackage[T1]{fontenc}
\usepackage[latin9]{inputenc}
\usepackage{color}
\usepackage{amsmath}
\usepackage{amssymb}
\usepackage{graphicx}
\usepackage[pdftex,unicode=true,pdfusetitle,
 bookmarks=true,bookmarksnumbered=false,bookmarksopen=false,
 breaklinks=false,pdfborder={0 0 1},backref=false,colorlinks=false]
 {hyperref}

\makeatletter

\begin{document}
\title{Computational Materials 
Discovery for Lanthanide Hydrides at high pressure: predicting High Temperature superconductivity}
\author{Evgeny Plekhanov}
\affiliation{King's College London, Theory and Simulation of Condensed Matter
(TSCM), The Strand, London WC2R 2LS, UK}
\author{Zelong Zhao}
\affiliation{King's College London, Theory and Simulation of Condensed Matter
(TSCM), The Strand, London WC2R 2LS, UK}
\author{Francesco Macheda}
\affiliation{King's College London, Theory and Simulation of Condensed Matter
(TSCM), The Strand, London WC2R 2LS, UK}
\author{Yao Wei}
\affiliation{King's College London, Theory and Simulation of Condensed Matter
(TSCM), The Strand, London WC2R 2LS, UK}
\author{Nicola Bonini}
\affiliation{King's College London, Theory and Simulation of Condensed Matter
(TSCM), The Strand, London WC2R 2LS, UK}
\author{Cedric Weber}
\affiliation{King's College London, Theory and Simulation of Condensed Matter
(TSCM), The Strand, London WC2R 2LS, UK}

\date{\today}

\begin{abstract}
Hydrogen-rich superhydrides are believed to be very promising high-T$_c$ superconductors,
with experimentally observed critical temperatures near room temperature,
as shown in recently discovered lanthanide superhydrides at very high pressures, e.g. LaH$_{10}$ at 170 GPa and CeH$_9$ at 150 GPa. With the motivation of discovering new hydrogen-rich high-T$_c$ superconductors at lowest possible pressure, quantitative theoretical predictions are needed. In these promising compounds, superconductivity is mediated by the highly energetic lattice vibrations associated with hydrogen and their interplay with the electronic structure, requiring fine descriptions of the electronic properties, notoriously challenging for correlated $f$ systems. In this work, 
we propose a first-principles calculation platform with the inclusion of many-body corrections to evaluate the detailed physical properties of the Ce-H system and to understand the structure, stability and superconductivity of CeH$_9$ at high pressure. We report how the prediction of T$_c$ is affected by the hierarchy of many-body corrections, 
and obtain a compelling increase of T$_c$ at the highest level of theory, which goes in the direction of experimental observations.
 Our findings shed a significant light on the search for superhydrides in close similarity with atomic hydrogen within a feasible pressure range. We provide
 a practical platform to further investigate and understand conventional superconductivity in hydrogen rich superhydrides.
\end{abstract}

\maketitle

\section{Introduction}

Pressure, like temperature, is a basic thermodynamic quantity which can be applied in experiments over an enormous range, leading to important contributions in such diverse areas of science and technology as astrophysics, geophysics, condensed matter physics, chemistry, biology 
\cite{schilling2000use}. Pressure can lead to a variety of new phenomena, a striking example being Alkali metals, which exhibit a series of electronic phase transitions to superconducting \cite{matsuoka2009direct}, complex structural forms that might lack periodicity \cite{mcmahon2006composite}, and with peculiar optical reflectivity \cite{guillaume2011cold}, when compressed even modestly. 

In particular, the search for superconducting metallic hydrogen at very high pressures has long been viewed as a key problem in physics \cite{guillaume2011cold} and is not new. Ashcroft himself proposed very early that hydrogen-based materials containing other main group atoms might exhibit superconductivity at large temperature \cite{ashcroft2004hydrogen}. These considerations are based on the Bardeen-Cooper-Schrieffer (BCS) theory, where superconductivity is phonon-mediated, and has motivated many theoretical and experimental efforts in the search for high-temperature superconductivity in hydrides at high pressures \cite{Eremets1506}. A recent milestone has been achieved with (near-) room temperature superconductivity in hydrogen disulfide \cite{drozdov2015conventional,snider2020room}. 

Hydrogen disulfide has not previously been considered as a superconductor because it was believed to go through dissociation at high pressure. Recent theoretical work~\cite{li2014metallization} indicated however that the dissociation would not occur, and Li and co-workers predicted that the material would become superconducting at $1.6$ million atmospheres, with temperatures above $80$ K. This led to the practical experimental work of Drozdov et al. \cite{drozdov2015conventional}, who compressed the hydrogen disulfide in a diamond anvil cell. Theoretical calculations predicted that hydrogen sulfide would transform on compression to a superconductor with a T$_c$ up to $200$ K \cite{li2014metallization}. The high T$_c$ of $203$ K at $150$ GPa in samples formed by compression of H$_2$S was subsequently confirmed~\cite{einaga2016crystal}.

This discovery has motivated scientists to expand further the scope of research to lanthanide hydrides under pressure, such as La-H \cite{kostrzewa2020lah} and Y-H \cite{li2014metallization}.  For the La-H systems, remarkably, not only the d electrons of La and $s$ electrons of H contribute to the Fermi density and hence to superconductivity, but also the $f$ electrons of the lanthanide. 

Interestingly, DFT calculations have shown that the fcc LaH$_{10}$ is a good metal with several bands crossing the Fermi level along many directions, associated with a large electronic density at the Fermi level \cite{liu2017potential}. It was suggested that the  contribution of $f$ states is dominant, realizing a unique high-T$_c$ with $f$ electrons at the Fermi level, in sharp contrast with the YH$_{10}$ system, where only the d electrons of Y and s electrons of H are the main contributors to Fermi density at high pressures \cite{li2015pressure}. This is due to the fact that external pressure destabilizes La-$6s$ and La-$5d$ orbitals to a greater extent than La-$4f$, which populates upon pressure. This opens new exciting avenues into the prediction and design of novel superconductors at high pressure, which would involve $f$ electrons, notoriously difficult to model in DFT, due to the strong electronic correlation effects. 

Indeed, the effects of electron localisation and hybridisation are applicable to all high-density matter, but are in particular relevant for the large family of correlated materials, based on transition metal $d$ or lanthanide and rare-earth $f$ elements. Correlated systems show a breadth of peculiar and interesting properties stemming from many-body effects, such as high temperature superconductivity in copper oxides and room temperature metal-insulator transition in vanadates. A fair understanding has been obtained in particular with the Zaanen-Sawatsky-Allen (ZSA) theory~[\onlinecite{zaanen1985band}], which provides a classification of transition metal periodic solids in terms of charge transfer or Mott systems, but the understanding of their properties far from ambient conditions remains lacking and challenging.

Hence, with the motivation of discovering new hydrogen-rich high-T$_c$ superconductors at the lowest possible pressure, quantitative theoretical predictions are needed. In these promising compounds, superconductivity is mediated by the interaction between the highly energetic lattice vibrations of Hydrogen and electrons, requiring fine descriptions of the electronic properties, notoriously challenging for correlated $f$ systems.
Electronic correlations provide corrections for several theoretical
elements, in particular the electron-phonon coupling strength $\lambda$,  the phonon dispersion relations, the electron spectral weight, and cross terms between electron-electron and electron-phonon interactions. 

As many-body corrections to phonon-dispersion relations are generally less dramatic than corrections to spectral functions (see for instance Ref.~\onlinecite{dmft_phonons}), and their full treatment is beyond reach, we propose in this work a pragmatic first-principles calculation platform which provides many-body corrections for the electronic spectral weight, which in turns provides corrected estimates of $\lambda$ and superconducting temperatures. Indeed, many-body corrections in f systems can induce dramatic changes, with energy shifts on the scale of one electron-volt or more, albeit with a systematic improvement obtained by DMFT as observed in metal-oxygen in actinides and rare-earth oxides \cite{dmft_gaps}. We evaluate here the detailed physical properties of the Ce-H system, and we focus in particular on the effect
of spectral transfer induced by electronic correlations. 

CeH$_9$ crystallizes in a P6(3)/mmc clathrate structure with a very dense three-dimensional atomic hydrogen sublattice at 100 GPa.
We report how the prediction of T$_c$ is affected by the hierarchy of many-body corrections, and obtain a remarkable agreement at the highest level of theory. 
 Our findings shed a significant light on the search for superhydrides in close similarity with atomic hydrogen within a feasible pressure range. We provide
 a practical platform to further investigate and understand conventional superconductivity in hydrogen rich superhydrides.

\section{Discussion}

The idea that hydrogen-rich compounds may be potential high-T$_c$ superconductors has originated from the beginning of the millennium, when chemical pre-compression was proposed as an effective way to reduce the metallization pressure of hydrogen by the presence of other elements, leading to observed T$_c$ exceeding 200 K in LaH$_{10}$ system \cite{lah} and 130 K in CeH$_9$ \cite{ceh} have indicated compressed hydrogen-rich compounds as potential room-temperature superconductors. 

It is recognized that superconductivity in such hydrides owes its origin to electron-phonon coupling. According to BCS theory, $T_c$ is mainly determined by four parameters: the characteristic phonon frequency, the electron-phonon coupling strength, the density of states at Fermi level, and the $\mu^*$. It is widely accepted that density functional theory with standard pseudo-potentials, such as PBE, provides an accurate description of the lattice dynamics. However, DFT is notoriously struggling with the treatment of strong electronic correlations for compounds with weakly hybridised and localised f electrons, such as Ce, where many-body corrections are called for. In this work, we use the density functional theory approach extended with dynamical mean-field theory (DMFT). DMFT readily provides corrections associated with the charge and spin local fluctuations, relevant in particular for the local paramagnetic moment of lanthanide elements. Typically, DMFT accounts for changes of orbital character at the Fermi surface, due to spectral weight transfer associated with the splitting of Hubbard f bands. As described in the method section, this affects in turn low energy electron-electron scattering processes via phonon momentum transfer, as captured by the Allen Dynes formalism.  
We focus hereafter on the CeH$_9$ P6(3)/mmc system at 200 GPa.  

\begin{figure*}
\includegraphics[width=2\columnwidth]{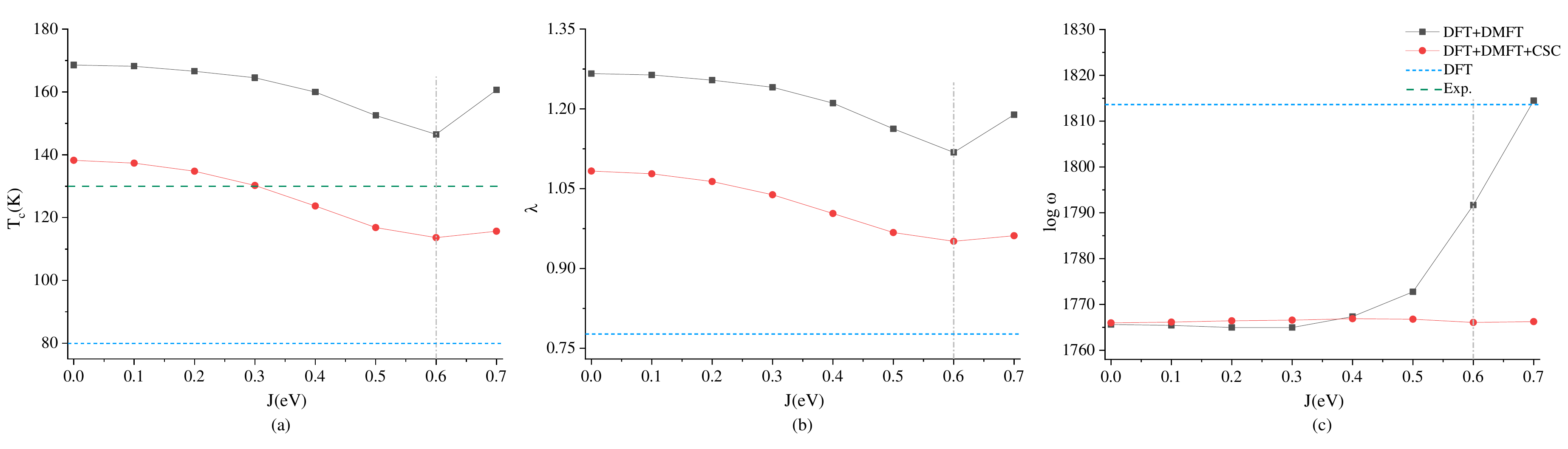}
\caption{\textbf{Many-body corrections to the superconducting temperature}. We report (a) the superconducting temperature $T_{c}$ obtained by the Allen and Dynes formalism, (b) the electron-phonon coupling strength $\lambda$ and (c) $\log(\omega)$, as a function of the Hund's coupling J. The spectral weight at the Fermi level is obtained at different levels of approximation: i) DFT PBE (horizontal light blue short-dashed line), ii) with many-body corrections obtained by one-shot dynamical mean-field theory (DFT+DMFT, filled black squares), and iii) with the fully charge self-consistent formalism (DFT+DMFT+CSC, filled red circles). 
The many-body corrections systematically improve the DFT T$_c$ in the direction of the experimental
value (horizontal long-dashed line). The one-shot DMFT provides a large increase of the superconducting temperature, overshooting largely the experimental value, which is concomitant with a sizeable increase of $\log(\omega)$. The charge self-consistency mitigates this effect--- $\log(\omega)$ remains in-line with the PBE value--- and in turns reduces the T$_c$. The physical value of the Hund's coupling for Ce ($J\approx 0.6$ eV) is reported by the vertical dashed line in all panels. All calculations were performed in the P6(3)/mmc phase of CeH$_9$ at 200 GPa.
}
\label{fig:yao1}
\end{figure*}

Firstly, we explore the effects of the many body corrections at various levels of theory (see Fig.\ref{fig:yao1}). We observe that the predicted superconducting temperature $T_{c}$ obtained by the Allen and Dynes formalism is largely affected by correlation effects. In particular, we compare: i) PBE density functional theory, ii) DFT correction with many-body corrections obtained by one-shot dynamical mean-field theory (DFT+DMFT), and iii) DFT+DMFT with the full charge self-consistent formalism (DFT+DMFT+CSC). We used the typical Koster-Slater
interaction vertex for the Ce correlated many-fold, with typical values for Ce ($U=6$eV, $J=0.6$eV). As the Hund's coupling controls the splitting of the unoccupied Ce f-states spectra feature into magnetic multiplets, we provide in Fig. \ref{fig:yao1} the dependence of $T_{c}$ on $J$. 
Interestingly, the one-shot DMFT provides a large increase of the superconducting temperature (see panel a), largely overshooting the experimental value $T_c=130$ K, which is concomittant with a sizeable increase of $\log(\omega)$. The charge self-consistency mitigates this effect ($\log(\omega)$ remains in-line with the PBE value), and in turns reduces the T$_c$. 
This confirms that many-body effects have a sizeable contribution to the prediction of the superconducting temperatures in lanthanide hydrides. Note that we use $J=0.6$ eV though out the rest of the paper.

\begin{figure}
\includegraphics[width=1\columnwidth]{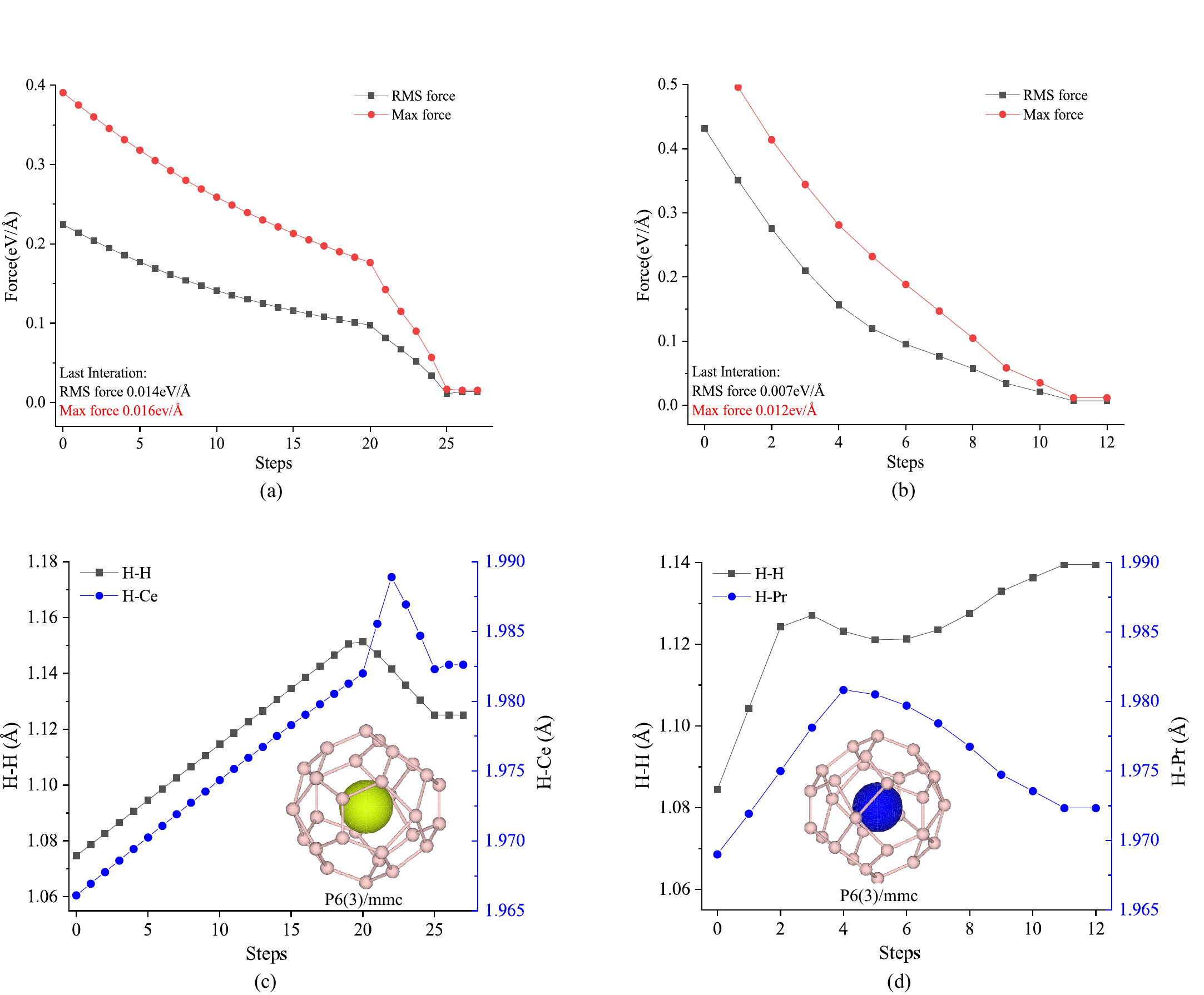}
\caption{\textbf{Structural relaxation of clathrate lanthanides with many-body corrections}. 
We report the structural relaxation of the CeH$_9$ (panel a and c) and prototype PrH$_9$ (panel b and d) compounds. All calculations are performed at $200$ GPa. The volume density is obtained by the equation of state in DFT+DMFT+CSC that provides very similar results to PBE (not shown). Internal coordinates are relaxed with DFT+DMFT+CSC, building upon the recent implementation of DFT forces for ultra-soft pseudo-potentials \cite{evgeny_ultrasoft}. 
We report the forces and total energies obtained during the structural optimisation, respectively, in panel a (b) for CeH$_9$ (PrH$_9$). Convergence is obtained within $25$ iterations. The shortest H-H and Ce-H bond lengths increase throughout the structural optimisation
(see panel c and d for Ce and Pr hydrides, respectively).}
\label{fig:yao2}
\end{figure}

Although DMFT readily provides important corrections to the electronic character, it is 
worthwhile to explore its effect on the structural properties. As the computational
overhead to perform many-body corrections remains significant, calculating forces
with finite atomic displacement is not tractable. Recent progress has however been made in this direction ~\citep{Haule2016}, in particular with the generalisation of the Hellmann-Feynmann theorem for 
DMFT in presence of ultra-soft pseudo-potentials \cite{evgeny_ultrasoft} in CASTEP \citep{CASTEPDMFT}. This generalisation opens new avenues for systems with heavy elements, not well suited for all-electron calculations. We report in Fig. \ref{fig:yao2} the structural
relaxation at $200$ GPa. We typically obtain corrections for the bond lengths of order 4$\%$. As expected, many-body effects tend to increase the length of Ce-H bonds, associated with a reduction of the hybridisation induced by electronic correlations. This bond length increase is also concomitant with a small increase of the minimum H-H distance. 
The volumetric density is also weakly affected by many-body effects, as the pressure obtained from  DFT+DMFT+CSC calculations on the DFT structure yields $216$ GPa. We hence conclude that DFT is a reasonable approximation for the structural properties of CeH$_9$. 

\begin{figure*}
\includegraphics[width=2\columnwidth]{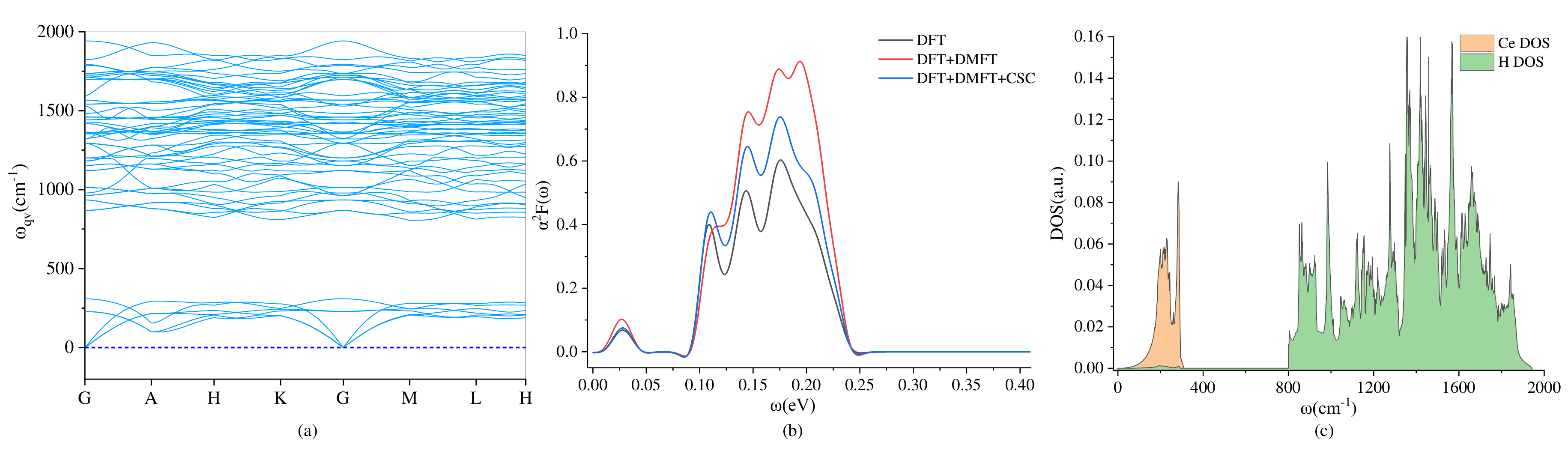}
\caption{\textbf{Lattice dynamics of CeH$_9$ at high pressure.}(a) Phonon dispersion relation, (b) Eliashberg function $\alpha^2F(\omega)$ and (c) phonon density of states for the P6(3)/mmc phase of CeH$_9$ at $200$ GPa. The phonon density of states is resolved in the Ce and H contributions. The dominant contribution to the Eliashberg function is due to the hydrogen vibrational modes located above the phonon gap ($\omega>750$cm$^{-1}$).}
\label{fig:yao3}
\end{figure*} 

The lattice dynamics for CeH$_9$ (see Fig. \ref{fig:yao3}) leads to a typical phonon gap
for lanthanide clathrates (between $\approx 300$ to $750$cm$^{-1}$), and high frequency modes
dominated by Hydrogen character ($\approx 750$ to $2000$cm$^{-1}$). The latter leads to 
a large weight in the Eliashberg function between $0.1$ to $0.25$eV, i.e. in the region that mostly contributes to the electron-phonon coupling strength $\lambda$. The effect of many-body corrections is indicated
in Fig. \ref{fig:yao3}.b, with an increase in the latter energy region due to the DMFT corrections (the trend follows the one observed in Fig \ref{fig:yao1}.a). 

\begin{figure}
\includegraphics[width=1\columnwidth]{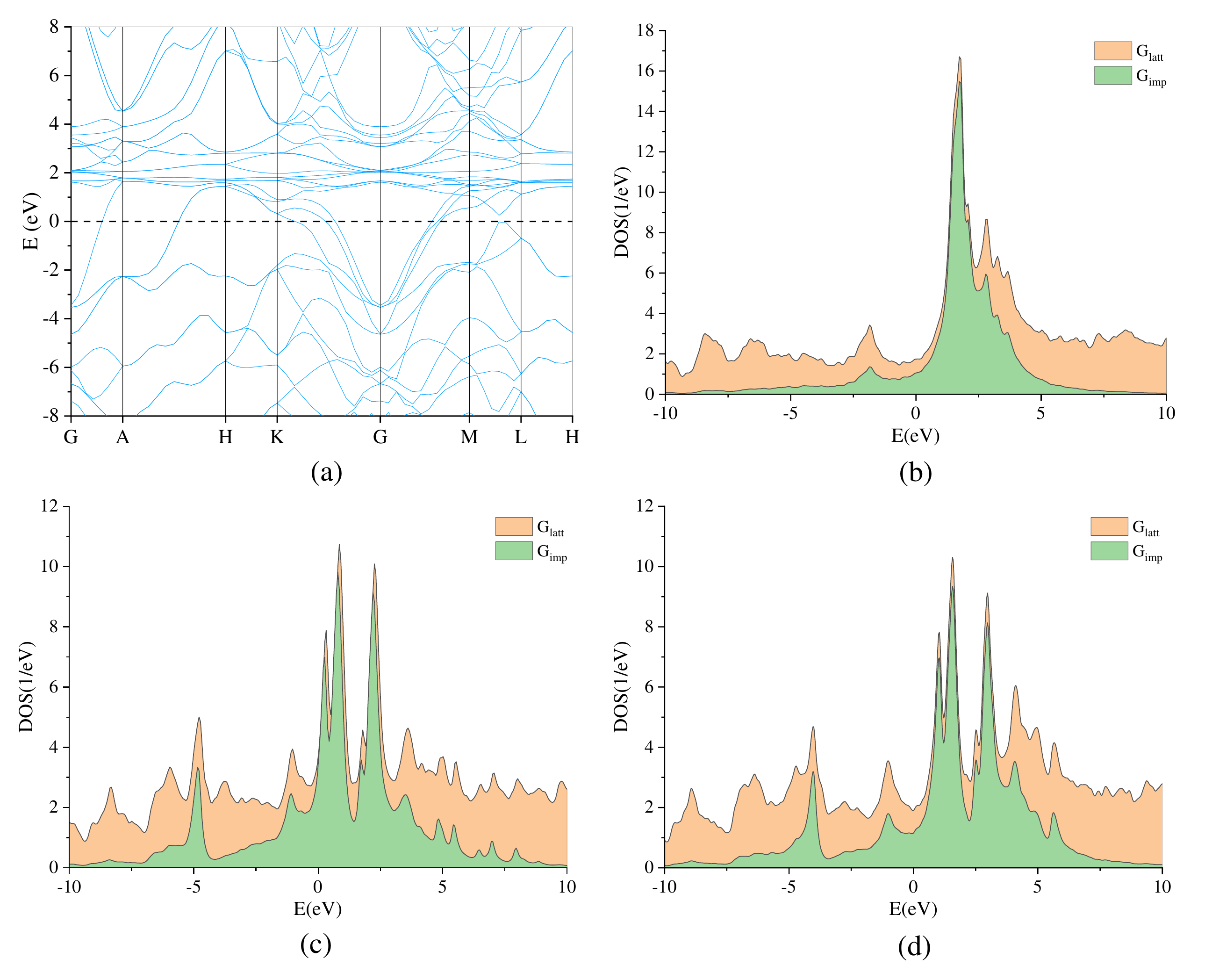}
\caption{\textbf{Spectral weight transfer induced by many-body corrections}.(a) Electronic band structure and (b) density of states obtained by DFT calculations. $G_{lat}$ and $G_{imp}$ denote the spectral weight obtained by the imaginary part of respectively the lattice and f impurity Green's function, corresponding to the spectral weight traced over all orbitals and traced over the f orbitals, respectively. In (c) and (d) we show the energy-resolved spectral weight, obtained respectively by the one-shot DFT+DMFT and the full charge self-consistent DFT+DMFT+CSC. All calculations were performed in the P6(3)/mmc phase of CeH$_9$ at 200 GPa.}
\label{fig:yao4}
\end{figure}

The changes highlighted above stem directly from a spectral weight transfer induced by many-body corrections (see Fig. \ref{fig:yao4}.a). In DFT, the Ce system is described by a two band system in absence of long-range magnetic order. We note that DFT is a single Slater determinant approach, and hence can not capture the role of paramagnetism, with an associated
magnetic multiplet (fluctuating magnetic moment). Such effects typically induce a splitting
of spectral features into satellites, as observed in Figs. \ref{fig:yao4}.b,c, with a resulting large increase of f-character at the Fermi level. As sharp Ce features occur near the Fermi level, we emphasize that a high level of theory is required to capture correctly the superconducting properties. For instance, in our calculations the full charge self-consistent approach (DFT+DMFT+CSC) induces a small shift of the sharp Ce feature at the Fermi level, which
in turns mitigates the f character increase at the Fermi level. 

\begin{figure}
\includegraphics[width=1\columnwidth]{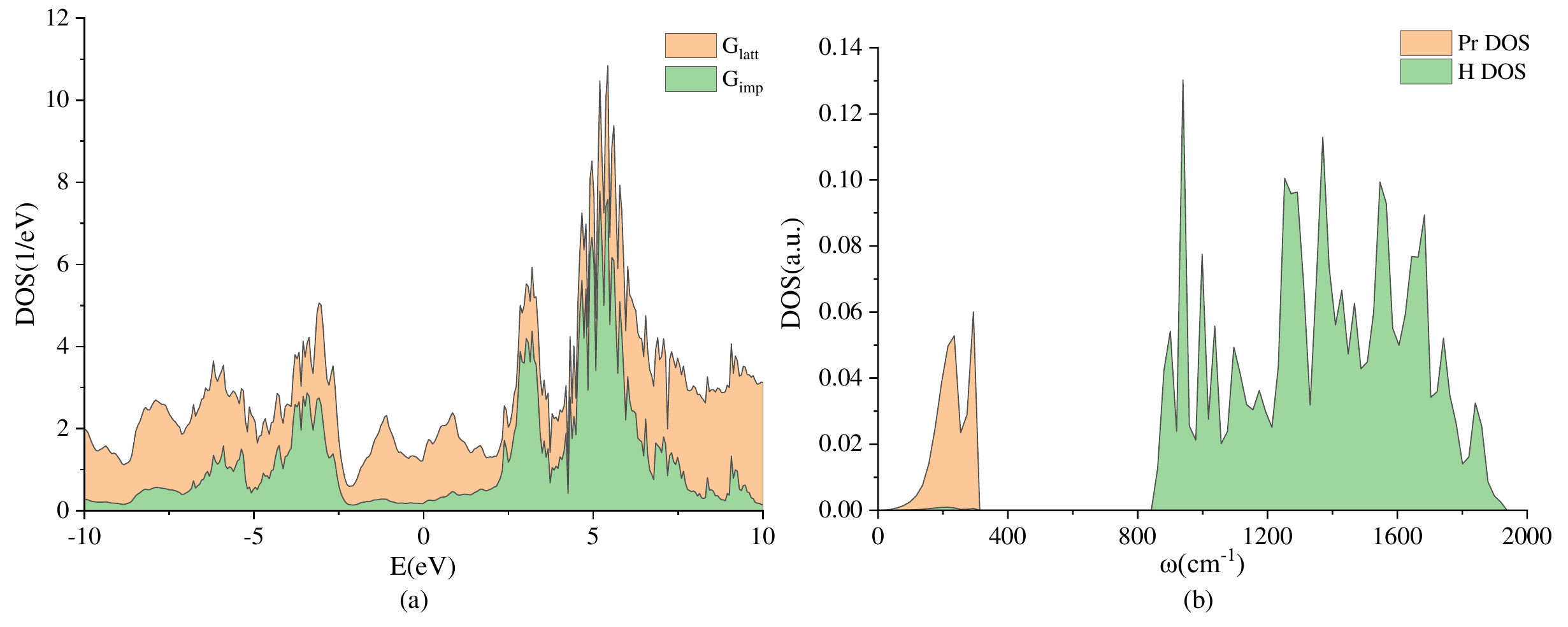}
\caption{\textbf{Aliovalent clathrate Pr prototype}. (a) Electronic density of states obtained by DFT+DMFT+CSC and (b) phonon density of states for the P3(6)/mmc phase of PrH$_9$ at $200$ GPa; the density of states is resolved in the Pr and H contributions. In stark contrast with the CeH$_9$ clathrate, the spectral weight of f electrons is suppressed at the Fermi level, resulting in a lower $T_c$.}
\label{fig:yao5}
\end{figure}

As the role of $f$ electronic orbitals is paramount for the superconducting properties, we study a
prototype lanthanide clathrates with higher $f$ occupations, by considering the 
aliovalent praseodymium hydride PrH$_9$. We note that this system is only metastable
at $200$ GPa in experiments \cite{pr_hydride}, although stable in the pressure range
$90-140$ GPa with a superconducting temperature $T_c=55$K above $110$ GPa.
We report in Fig. \ref{fig:yao5} the DFT+DMFT+CSC framework applied to PrH$_9$ in the P6(3)/mmc phase at 200 GPa. We obtain a theoretical estimate of $T_c=70$K, in qualitative agreement
with the experimental value obtained at lower pressure, where the P6(3)/mmc phase is stable.
In line with the CeH$_9$ calculations, DFT also underestimates the critical temperature
of PrH$_9$, with $T_c=12$K for $\mu^\star=0.1$. 

We attribute the decrease in $T_c$ with the higher f-occupation, which shifts the chemical
potential away from the f- spectral features present near the Fermi level (see Fig. \ref{fig:yao5}.a). As expected, the lattice dynamics (see Fig. \ref{fig:yao5}.b) are qualitatively similar to the CeH$_9$ compound. 

Our results highlight a pathway for the optimisation of $T_c$ in lanthanide hydrides: the increase of $f$ character at the Fermi level in DFT+DMFT, associated with a smaller degree of Ce-H covalency and lesser degree of hybridisation, provides a marker for an increased superconducting temperature in these systems.

\section{Methods}

\begin{figure}
\includegraphics[width=1\columnwidth]{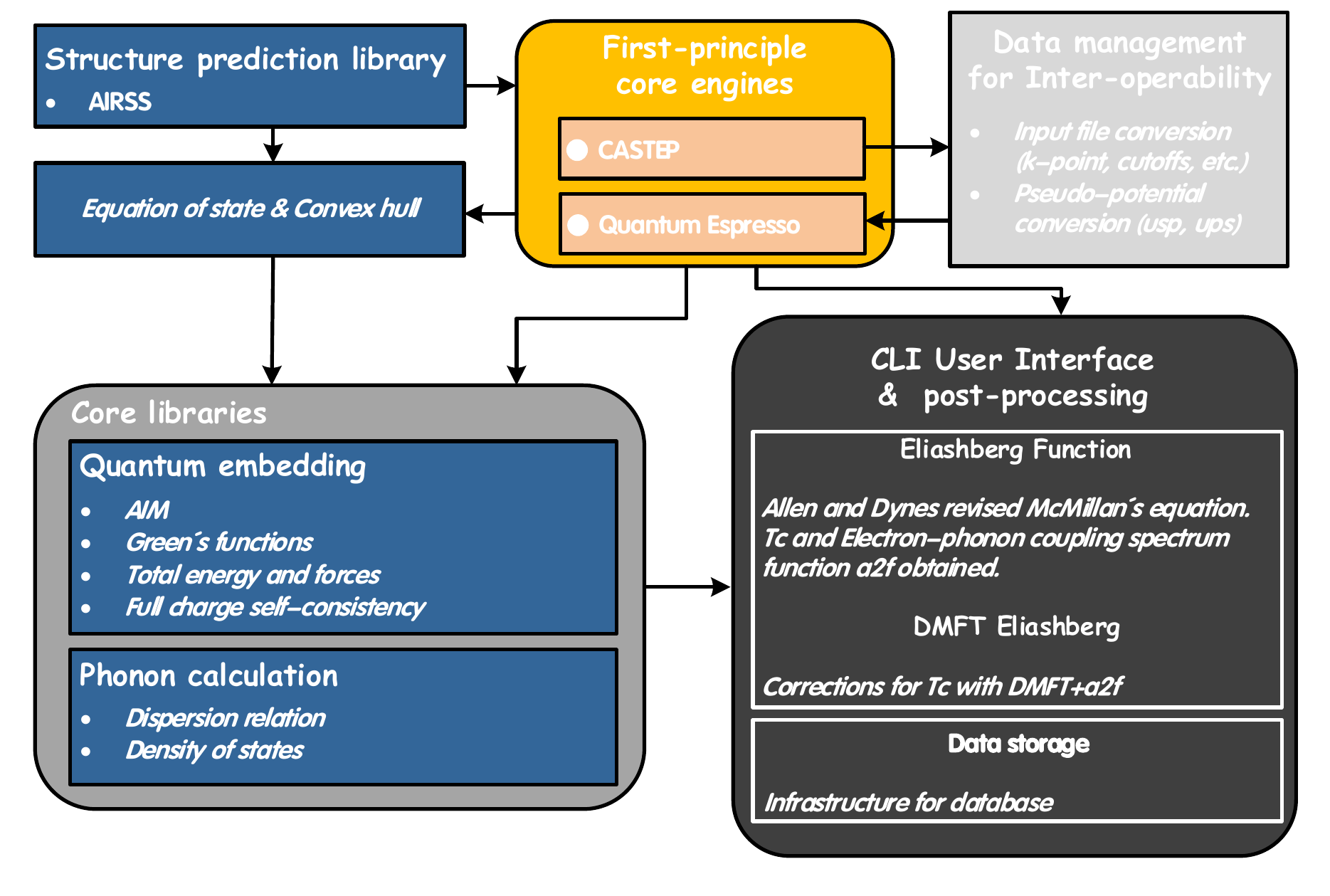}
\caption{\textbf{DFT inter-operability for a consistent many-body platform}. Schematic overview of the main modules of the theoretical platform and its interrelations. Firstly, structures are predicted by Ab Initio Random Structure Search (AIRSS), via Gibbs enthalpies for the equation of state and convex hull. The underlying core engines are the CASTEP and Quantum Espresso DFT softwares. Inter-operability between QE and CASTEP is achieved via format conversion of input files, pseudo-potentials and $\mathbf{k}$-point grids. 
Core libraries are used to provide the many-body corrections, via quantum embedding, which in turns provides corrected forces and energies. In post-processing, the Eliashberg function and superconducting T$_c$ are obtained with the DMFT+a2F approach. Finally, data are archived for future usage in hdf5.}
\label{fig:workflow1}
\end{figure}

We present our theoretical approach in Fig. \ref{fig:workflow1}. We
report a schematic overview of the main modules of the theoretical platform and their interrelations. Our approach provides a modular framework for systematic material screening
at high pressure. Firstly, stochiometric compositions are provided by Ab Initio Random Structure Search (AIRSS) \cite{airss}, via Gibbs enthalpies for the equation of state and convex hull. The underlying core engines are CASTEP \cite{CASTEP} and Quantum Espresso \cite{QE_2009}, both interfaced with AIRSS. 
Inter-operability between QE and CASTEP is achieved via format conversion of input files, pseudo-potentials and k-point grids. Core libraries are used to provide the many-body corrections, via the DMFT quantum embedding, which in turns provides corrected forces \cite{evgeny_ultrasoft} and total energies \cite{CASTEPDMFT,Lee2019}. 

The underlying structure relaxations were carried out using the QE and CASTEP packages in the framework of DFT and using PBE-GGA (Perdew-Burke-Ernzerhof generalized gradient approximation)~\cite{perdew1992atoms,perdew1996generalized}.
Ultra-soft pseudo-potentials were used to describe the core electrons and their effects on valence orbitals. Valence electron configuration of $5s^2 5p^6 4f^1 5d^1 6s^2$ (i.e., with explicitly included $f$ electrons) and $1s^1$ was used for the Ce and H atoms, respectively. A plane-wave kinetic-energy cut-off of $1000$ eV  and dense Monkhorst-Pack k-points grids with reciprocal space resolution of $2\pi \times 0.07$ \AA$^{-1}$ were employed in the calculation. 

Phonon frequencies and superconducting critical temperature were calculated using density-functional perturbation theory as implemented in the Quantum Espresso. The $\mathbf{k}$-space integration (electrons) was approximated by a summation over a $24\times 24\times 12$ uniform grid in reciprocal space, with the smearing scheme of Methfessel-Paxton using a temperature $k_B$T = 0.05 eV for self-consistent cycles and relaxations; the same grid ($24\times 24 \times 12$) was used for evaluating DOS and coupling strength. Dynamical matrices and $\lambda$ were calculated on a uniform $6\times 6\times 3$ grid in $\mathbf{q}$-space for P63/mmc-CeH$_9$. 

In post-processing, the superconducting transition temperature $T_c$ was estimated using the Allen-Dynes modified McMillan equation:
 
\begin{equation}
T_c = \frac{\omega_{\log}}{1.2} \exp \left [ 
         \frac{-1.04(1+\lambda)}{\lambda - \mu^\star(1+0.62\lambda)}\right ]
\end{equation}
where $\mu^{*}$ is the Coulomb pseudopotential. The electron-phonon coupling strength $\lambda$ and $\omega_{\log}$ were calculated as:
\begin{equation}
\omega_{\log} = \mbox{exp} \left [ \frac{2}{\lambda} \int \frac{d\omega}{\omega}
                                  \alpha^2F(\omega) \log(\omega) \right ],
\end{equation}

\begin{equation}
\lambda = \sum_{\mathbf{q}\nu} \lambda_{\mathbf{q}\nu} = 
2 \int \frac{\alpha^2F(\omega)}{\omega} d\omega.
\end{equation}

In the Allen-Dynes formalism, the Eliashberg function $\alpha^2F(\omega)$ is obtained by summing over all scattering processes at the Fermi level mediated by phonon momentum transfer and reads \cite{allendynes}:

\begin{equation}
\alpha^2F(\omega) = N(\epsilon_F) \frac{\sum_{\mathbf{k_1},\mathbf{k_2}}{
                    \left|
                    M_{\mathbf{k_1},\mathbf{k_2}}
                    \right|^2
                    \delta(\omega-\omega_{\mathbf{q}\nu}) \delta(\epsilon_{\mathbf{k_1}}) \delta(\epsilon_{\mathbf{k_2}})}}
                    {\sum_{\mathbf{k_1},\mathbf{k_2}}{
                     \delta(\epsilon_{\mathbf{k_1}}) \delta(\epsilon_{\mathbf{k_2}})}}.
\end{equation}

Here, $N(\epsilon_F)$ is the DOS at Fermi level, $\omega_{\mathbf{q}\nu}$ is the phonon spectrum of a branch $\nu$ at momentum $\mathbf{q}=\mathbf{k_2}-\mathbf{k_1}$, $\epsilon_{\mathbf{k_1}}$ and $\epsilon_{\mathbf{k_2}}$ are electronic band energies referred to the Fermi level, while $M_{\mathbf{k_1},\mathbf{k_2}}$ are the electron-phonon coupling matrix elements. Many-body effects introduce a change of spectral character at the Fermi level, where electronic correlations induce a mass enhancement and introduce a finite lifetime, due to incoherence. In this spirit of the DMFT scissors \cite{TomczakBands},
we correct the DFT bands with the renormalised DMFT band picture:
\begin{equation}
\alpha^2F(\omega) = \mathcal{A}_{tot} \frac{\sum_{\mathbf{k_1},\mathbf{k_2}}{
                    \left|
                    M_{\mathbf{k_1},\mathbf{k_2}}
                    \right|^2
                    \delta(\omega-\omega_{\mathbf{q}\nu}) \mathcal{A}(\mathbf{k_1}) \mathcal{A}(\mathbf{k_2})
                    }}
                    {\sum_{\mathbf{k_1},\mathbf{k_2}}{
                     \mathcal{A}(\mathbf{k_1}) \mathcal{A}(\mathbf{k_2})}},
\end{equation}
where $\mathcal{A}_{tot}$ and $\mathcal{A}(\mathbf{k})$ are respectively the total and $\mathbf{k}$-momentum resolved spectral weights at Fermi level. This approach is
denoted as \textit{DMFT+a2F} in the workflow.

\begin{figure}
\includegraphics[width=1\columnwidth]{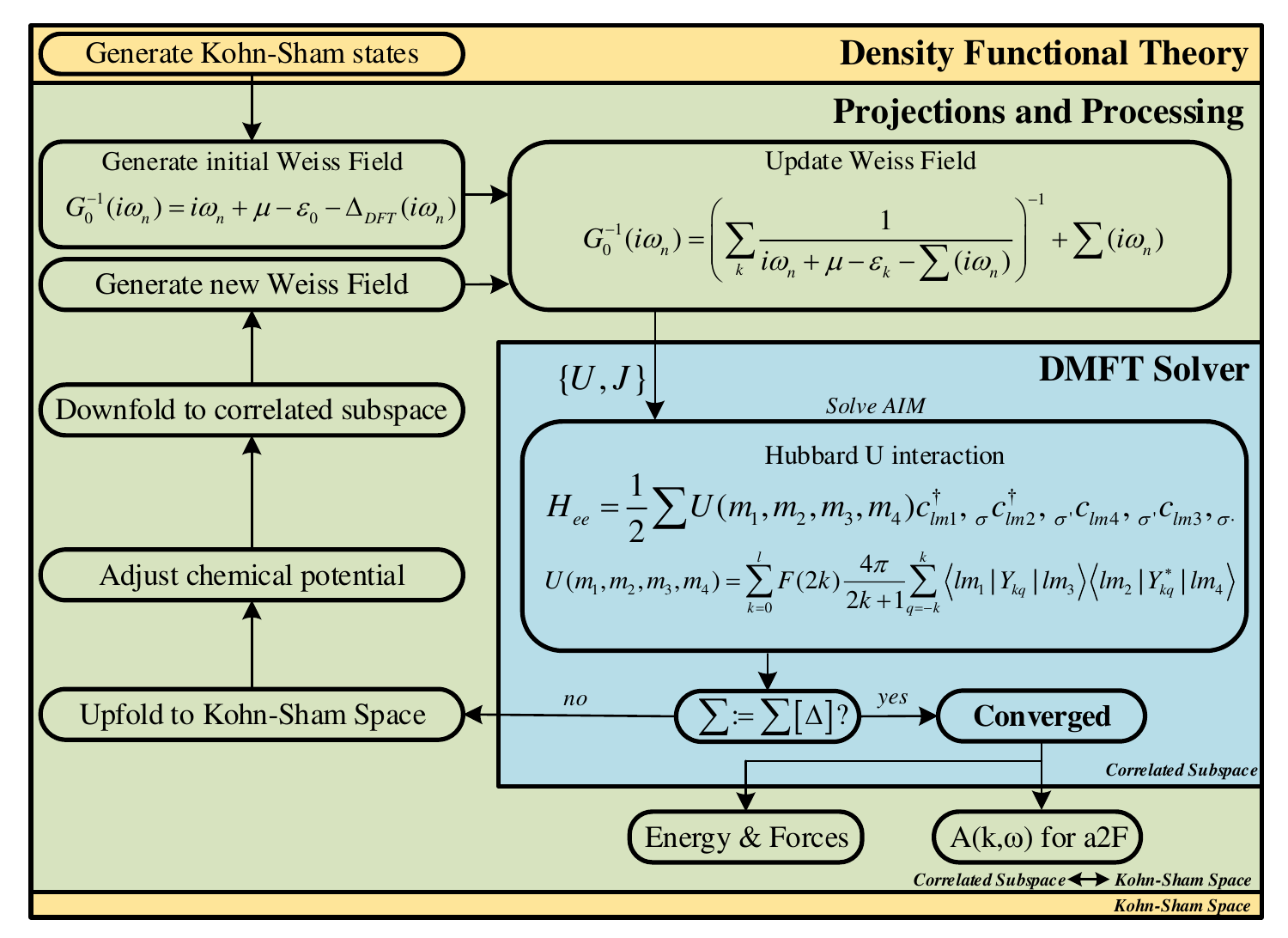}
\caption{\textbf{The fully charge-self consistent DMFT approach}. Flowchart of the DFT+DMFT calculations. The outer-region denotes the
density functional theory, which provides the Kohn-Sham potentials that in turn
feed into the calculation of the Green's function (mid-region). The projected
Green's function provides in turn the Weiss field, that is used to define
the Anderson Impurity Model, solved in the presence of a general two-body correlation term (inner-region). In the fully charge-self consistent approach (DFT+DMFT+CSC), the Kohn-Sham potentials are calculated from the DMFT electronic density. Upon DMFT convergence, total energies and forces are calculated from the Green's function and self-energy. The colours separate the procedure by levels of theory: the outer region deals with DFT, the mid-regions is dealing with the interface between DFT and the AIM, the inner region is where the AIM is solved by the quantum solver.}
\label{fig:workflow2}
\end{figure}

Fig. \ref{fig:workflow2} indicates the DMFT quantum embedding approach. 
Density functional theory provides the Kohn-Sham Hamiltonian that in turn
feed into the calculation of the DMFT Green's function. We use atomic projectors
to define the Anderson Impurity Model (AIM), solved with a Hubbard-I solver. 
A breadth of quantum solvers are readily available in the TRIQS open-source platform \cite{triqs1,triqs2}. In the fully charge-self consistent approach (DFT+DMFT+CSC), the Kohn-Sham potentials are calculated from the DMFT electronic density, obtained by the trace of Green's function \cite{CASTEPDMFT}. Upon DMFT convergence, total energies and forces are calculated from the Green's function and self-energy.

\section{Conclusion}

We have developed a framework that can readily 
provide many-body corrections to the estimation of the superconducting temperature
in lanthanide hydrides. We observe sizeable corrections to the uncorrected first-principles calculation when keeping in account many-body effects, and recover consistent agreement with experiments for CeH$_9$ at $200$ GPa. We report that the DMFT charge self-consistency, \textit{i.e.} the many-body corrections to the local charge density in the first-principles calculations, is instrumental to recover a consistent theoretical framework.
The increase of predicted superconducting temperature is induced by
a shift of the spectral weight of the $f$ states, which in turns affects the 
spectral character at the Fermi level. Albeit hitherto the many-body corrections remain limited to the electronic contribution to the Eliashberg function, we have
discussed the capabilities for relaxing lanthanide hydrides within the DMFT formalism,
building upon our recent developments that provide DMFT forces for DFT with norm-conserving and ultra-soft pseudo-potential. The latter provides structural insights, and we report that at large pressure many-body effects have small effects on the lattice, as DFT structures
and pressures are close to the DMFT counterparts. Although a full treatment of phonons at the DMFT level remains out-of-reach for such complex materials, the latter suggests that
the largest corrections to T$_c$ lies with many-body corrections to the electronic part.
We have studied the aliovalent effect, and observed that the increase of f-character at the Fermi level, as compared to an iso-structural Pr hydride, leads to an increase of superconducting temperature, a compelling observation for future explorations of f-systems as high T$_c$ superconductors.
Our approach is general and provides a modular framework to inter-operate typical first-principles software, i.e. the freely available CASTEP+DMFT code and Quantum Espresso, with a relatively small numerical footprint and is straightforward to implement. 
\quad

\section*{Data Availability}
The codes are available at url \textbf{dmft.ai} under the GPL 3.0 license. 
\quad

\section*{Acknowledgements}
CW, NB and EP are supported by the grant [EP/R02992X/1] from the UK Engineering and Physical Sciences Research Council (EPSRC). 
This work was performed using resources provided by the ARCHER UK National Supercomputing Service and the Cambridge Service for Data Driven Discovery (CSD3) operated by the University of Cambridge Research Computing Service (www.csd3.cam.ac.uk), provided by Dell EMC and Intel using Tier-2 funding from the Engineering and Physical Sciences Research Council (capital grant EP/P020259/1), and DiRAC funding from the Science and Technology Facilities Council (www.dirac.ac.uk).

\quad

\section*{Additional Information}
Correspondence should be addressed to Evgeny Plekhanov (evgeny.plekhanov@kcl.ac.uk), Nicola Bonini (nicola.bonini@kcl.ac.uk) and Cedric Weber (cedric.weber@kcl.ac.uk).

\bibliographystyle{apsrev}
\bibliography{yao,cedric}

\end{document}